\documentclass{emulateapj}

\usepackage{natbib}
\usepackage{url}

\shorttitle{QPP during the 15~February~2011 flare}
\shortauthors{Dolla et al.}

\begin{document}

\title{Time delays in quasi-periodic pulsations observed during the X2.2 solar flare on 15~February~2011}

\author{L.~Dolla\altaffilmark{1}, C.~Marqu\'e\altaffilmark{1}, D.~B.~Seaton\altaffilmark{1}, T.~Van Doorsselaere\altaffilmark{2,6}, M.~Dominique\altaffilmark{1}, D.~Berghmans\altaffilmark{1}, C.~Cabanas\altaffilmark{1}, A.~De~Groof\altaffilmark{1,3}, W.~Schmutz\altaffilmark{4}, A.~Verdini\altaffilmark{1}, M.~J.~West\altaffilmark{1}, J.~Zender\altaffilmark{3} and A.~N.~Zhukov\altaffilmark{1,5}}
\email{dolla@sidc.be}

\altaffiltext{1}{Solar-Terrestrial Center of Excellence, Royal Observatory of Belgium,
Avenue Circulaire 3, B-1180 Brussels, Belgium}
\altaffiltext{2}{Centrum voor Plasma-Astrofysica, Department of Mathematics, KULeuven, Celestijnenlaan 200B bus 2400, 3001 Leuven, Belgium}
\altaffiltext{3}{European Space Agency, ESTEC, Keplerlaan 1, 2201 AZ Noordwijk, The Netherlands}
\altaffiltext{4}{Physikalisch-Meteorologisches Observatorium Davos, World Radiation Center, Switzerland}
\altaffiltext{5}{Skobeltsyn Institute of Nuclear Physics, Moscow State University, 119992 Moscow, Russia}
\altaffiltext{6}{postdoctoral fellow of the FWO-Vlaanderen; funded by the European Community's seventh framework programme (FP7/2007-2013) under grant agreement number 276808. }

\begin{abstract}
We report observations of quasi-periodic pulsations (QPPs) during the X2.2 flare of 15~February~2011, observed simultaneously in several wavebands. We focus on fluctuations on time scale 1--30~s and find different time lags between different wavebands. During the impulsive phase, the Reuven Ramaty High Energy Solar Spectroscopic Imager (RHESSI) channels in the range 25--100~keV lead all the other channels. They are followed by the Nobeyama RadioPolarimeters at 9 and 17~GHz and the Extreme Ultra-Violet (EUV) channels of the Euv SpectroPhotometer (ESP) onboard the Solar Dynamic Observatory (SDO). The Zirconium and Aluminum filter channels of the Large Yield Radiometer (LYRA) onboard the Project for On-Board Autonomy (PROBA2) satellite and the SXR channel of ESP follow. The largest lags occur in observations from the Geostationary Operational Environmental Satellite (GOES), where the channel at 1--8~\AA{} leads the 0.5--4~\AA{} channel by several seconds. 
The time lags between the first and last channels is up to $\approx 9$~s.
We identified at least two distinct time intervals during the flare impulsive phase, during which the QPPs were associated with two different sources in the Nobeyama RadioHeliograph at 17~GHz. The radio as well as the hard X-ray channels showed different lags during these two intervals. 
To our knowledge, this is the first time that time lags are reported between EUV and SXR fluctuations on these time scales. We discuss possible emission mechanisms and interpretations, including flare electron trapping. 
\end{abstract}


\section{Introduction}
%
Quasi-periodic pulsations (QPPs) have been observed during solar flares for many years in many wavelengths, from radio to hard X-rays \citep{Nakariakov09} and even gamma rays \citep{Nakariakov10}\footnote{This identification is questioned by \citet{Gruber11}}. The observed periods range from fractions of a second to several minutes. Recent work reflects this diversity in time scales and wavelengths of observation: \citet{Foullon10, Kupriyanova10, Reznikova11}. 

\citet{Nakariakov09} divide the current interpretations into two classes: periodic load/unload mechanisms of non-thermal electrons produced during the flare (i.e. an intrinsic property of the reconnection mechanism that leads to a quasi-periodic behavior) or modulation of the electron beam or of parameters of the emitting plasma by magneto-hydrodynamic (MHD) waves. \citet{Fleishman08}, for example, interpret their observations in the framework of quasi-periodic injection of fast electrons, while for \citet{Asai01}, the injection is modulated by an oscillation of the flaring loop. \citet{Inglis08} favor an MHD wave, but do not exclude a quasi-periodic injection of the fast electrons. 

In this Letter we investigate this question by comparing the signal observed in different wavebands during a solar flare. We pay particular attention to the time delays and focus on short time scales fluctuations ($\lesssim 30$~s). 
%
\section{Observations}
%
Many instruments observed the X2.2 flare that started at 01:44~UT on 15~February~2011 in AR~11158 \citep{Schrijver11}. We focus here on the instruments that provide high time cadence (cf Table~\ref{tab observations}). 

The Large Yield Radiometer \citep[LYRA;][]{Hochedez06, Benmoussa09} on the Project for On-Board Autonomy (PROBA2) spacecraft observes at up to 100~Hz in four channels. We focus here on the Zirconium (Zr) and Aluminum (Al) channels, which both include contributions from soft X-rays (SXR) and Extreme Ultraviolet (EUV); see Table~\ref{tab observations} for details. LYRA's two UV channels, Lyman-$\alpha$ and Herzberg-continuum, contained no significant fluctuations, probably due to the degradation of the filters of the unit in use at the time of observation \citep{Dominique12}. 

The Nobeyama RadioPolarimeters (NoRP) measured for this event solar fluxes at 1, 2, 3, 9, 17 and 35~GHz. 
We also used reconstructed images from the Nobeyama RadioHeliograph (NoRH) at 17~GHz\footnote{available at \url{http://solar.nro.nao.ac.jp/norh/images/event/20110215_0154/steady_fujiki/}}. 

The Euv SpectroPhotometer \citep[ESP;][]{Didkovsky12} is part of the Extreme ultraviolet Variability Experiment \citep[EVE;][]{Woods12} instrument suite onboard the Solar Dynamic Observatory \citep[SDO;][]{Pesnell12}. It comprises five channels: one SXR and four EUV channels. 
The 18~nm waveband contains line emission that mainly comes from a plasma around 1--2~MK. 
The 26~nm and 30.4~nm channels mainly contain \ion{He}{2} lines. 
We subtracted an offset to correct an artificial jump in instrument bias that appeared beginning at 02:00~UT in all these channels. (We did not use EVE's 36~nm channel because of a technical problem with that channel.)

The Geostationary Operational Environmental Satellite (GOES) provides SXR fluxes in a ``short-wavelength'' bandpass (0.5--4~\AA{}) and a ``long-wavelength'' bandpass (1--8~\AA{}). 

The Reuven Ramaty High Energy Solar Spectroscopic Imager \citep[RHESSI;][]{Lin02} observes from 3 to 20000~keV. 
We restricted our analysis up to 300~keV, the emission being absent above that limit. 

We also used the Atmospheric Imaging Assembly \citep[AIA;][]{Lemen11} EUV imager onboard SDO for context identification. 
%
\section{Results}
%
\begin{figure*}
	\begin{center}
	\includegraphics[height=0.8\textheight]{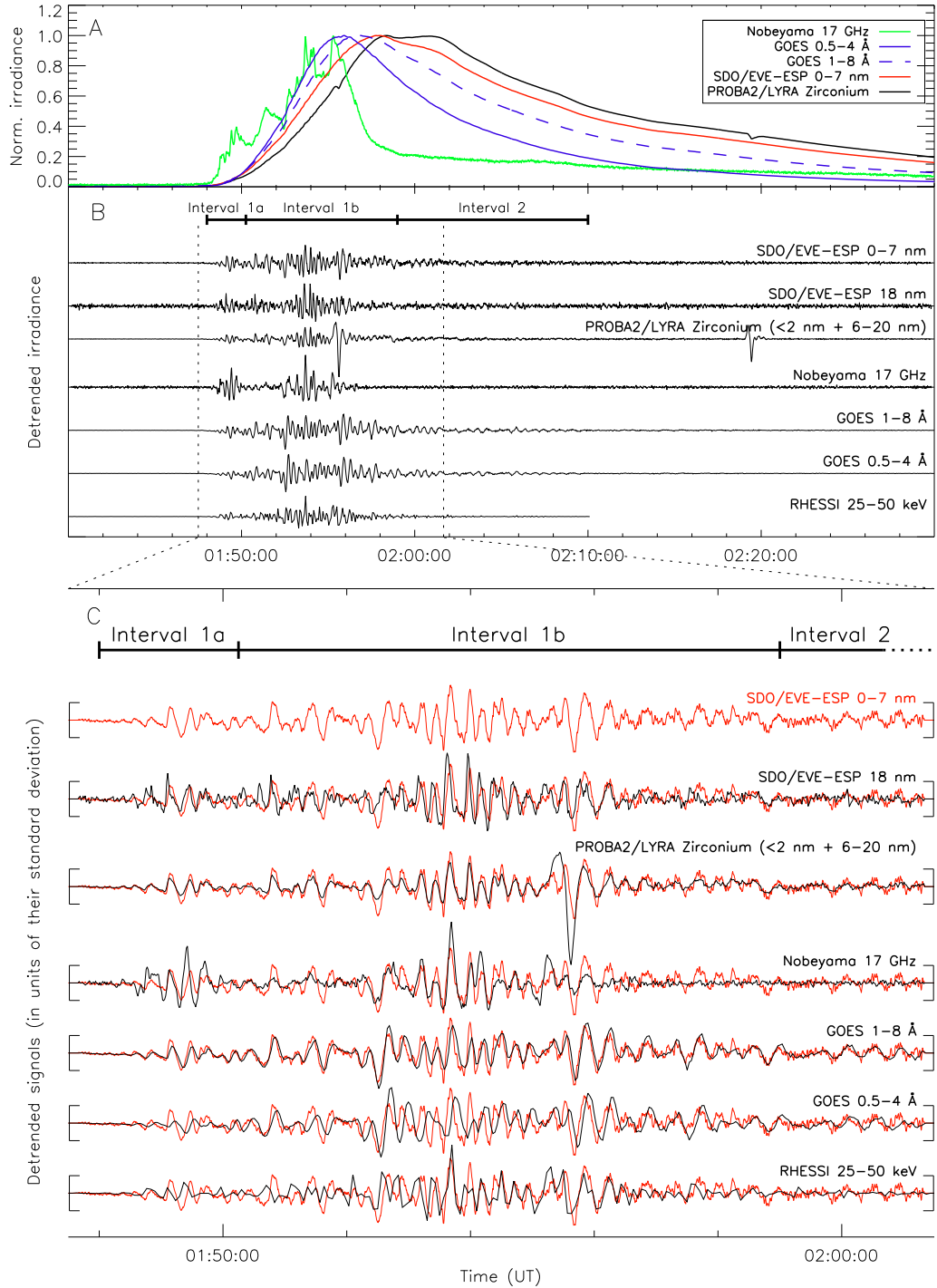}
	\end{center}
   \caption{Normalized irradiances in different instruments during the 15~February~2011 X2.2 flare (Panel~A). To emphasize the short-period fluctuations, Panel~B shows the same observations detrended by subtracting the signal smoothed using a 20-s boxcar (in units of their standard deviation). 
Panel~C shows a close-up view over the interval delimited by the vertical dotted lines in Panel~B. The ESP 0--7~nm light curve (red) is over-plotted on each curve for comparison. The tickmarks are drawn at $\pm 2$~standard deviations of the signals. 
The artifacts in the LYRA Zr curve around 01:56 and 02:20 are due to PROBA2 spacecraft maneuvers. 
}
	\label{fig light curves and detrended}
\end{figure*}
\begin{figure*}
	\begin{center}
	\includegraphics[height=0.95\textheight]{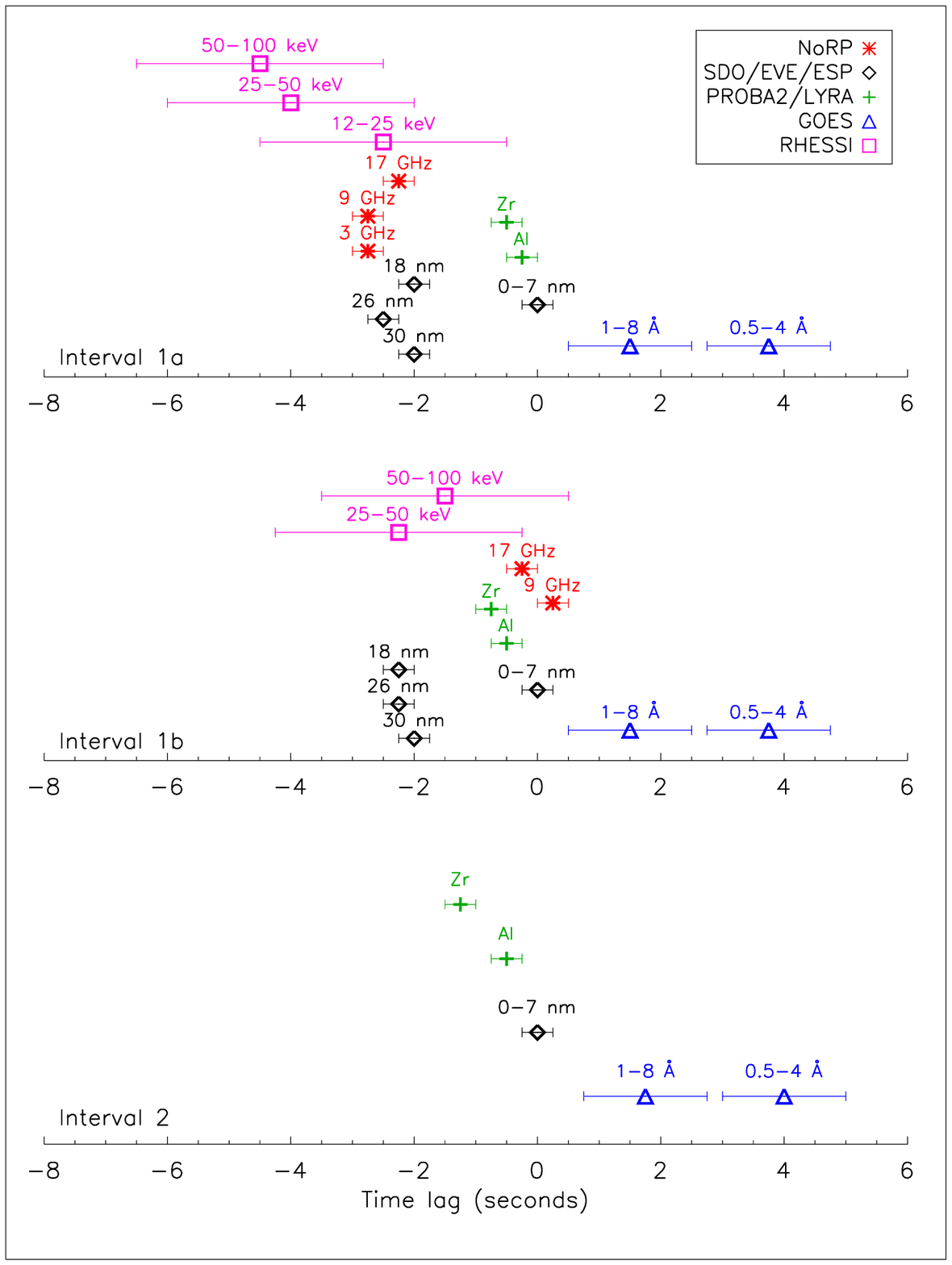}
	\end{center}
   \caption{Time delays for the detrended irradiances calculated with respect to the detrended irradiance of the ESP 0--7~nm channel (a negative delay means that the channel is leading the reference channel). Only channels with correlation coefficient larger than 0.4 are shown. Errors bars are based on the sampling time.  }
	\label{fig time lag}
\end{figure*}

Panel~A in Figure~\ref{fig light curves and detrended} presents the light curves we observed in several different channels.  The Nobeyama 17~GHz channel, like the RHESSI channel (not included), shows the impulsive phase of the flare. The other channels peak with different delays, depending on their respective waveband and temperature response. Panel~B shows several irradiance curves with the large-scale trend removed, which we achieve by subtracting the corresponding signal smoothed with a 20-s boxcar.  For simplicity we omit channels from each instrument that feature similar properties. 

Clearly, short time-scale fluctuations appear during the impulsive phase for each of the plotted channels. Some fluctuations of smaller amplitude are also present during the declining phase, for example, in the SXR channels.  In most cases, the fluctuations have a small amplitude relative to the overall increase in signal during the flare (Table~\ref{tab observations}). 

Panel~C shows an enlarged view where we superimpose every channel (black) over the ESP SXR channel (red). Although the curves of total irradiance have different shapes, especially the radio channel, the short term fluctuations ($\lesssim 20$~s) are strikingly similar. 

In most cases, time delays appear between similar features.  We investigated further by measuring the time delay that provides the maximum correlation between a channel and the ESP SXR channel, which we consider our reference channel; best correlation coefficients are shown in Table~\ref{tab observations}. We used linear interpolation to match the time cadence of the detrended signals (spline interpolation produced similar results).  We also smoothed each signal using a 1-s boxcar to reduce the effect of noise. This smoothing has a negligible effect, but the curves of correlation coefficient as a function of time lag are noisier and contain some spurious local maxima without it. Additionally, we verified that the best lags were not significantly different when using smoothing boxcars with widths from 5 to 30~s to remove the large-scale trend. We also cross-verified the optimal correlation lags by using several different channels for reference. 

It is clear that some local extrema in the RHESSI channels are nearly in anti-phase with those in ESP SXR, while some other extrema are in phase. As most extrema in the RHESSI channels from 3 to 100 keV coincide, the effect of Poisson noise can be ruled out. Nevertheless, it is difficult to distinguish between the effect of signal integration over the 4-s rotation of the spacecraft, instrumental noise, and real differences with the ESP SXR channel due, for example, to different contributing processes. Such differences could also be the result of the fact that soft and hard X-rays (HXR) originate in different parts of the flaring region.

Figure~\ref{fig time lag} synthesizes the delay times with respect to the ESP 0--7 nm channel for channels whose best correlation coefficients were larger than 0.4. 
There are three intervals over which the fluctuations exhibit clearly distinguishable behavior. Interval~1a lasts from 01:48:00 to 01:50:15; interval~1b lasts from 01:50:15 to 01:59:00. These intervals correspond to the early and late impulsive phase of the flare. Interval~2, from 01:59:00 to 02:10:00, corresponds to the portion of the declining phase where fluctuations are significant.

The main difference between intervals~1a and 1b is that some radio channels lead the reference channel during interval~1a, but do not during interval~1b. In fact, considering  possible offsets of the various instruments' clocks (believed to be $<1$~s), it is possible that the radio and ESP~0--7~nm channels are actually in phase during interval~1b. RHESSI time series also contain different lags during these two intervals as well. Thus we note that this behavior occurs particularly in wavebands characteristic of the impulsive phase. 

All the other channels retain roughly the same lags throughout the impulsive phase. The three EUV channels are nearly all in phase and lead all the SXR channels. Within the EUV and SXR channels we find, roughly, that harder spectral components contain larger lags. Additionally, both LYRA channels slightly lead the ESP reference channel, which is likely due to combination of EUV and SXR emission that contributes to these wavebands. 

During the declining phase (Interval~2), fluctuations appear in all SXR channels (but sometimes with correlation coefficient slightly smaller than 0.5), but not in the HXR, EUV or radio channels. 

\begin{figure*}
	\begin{center}
	\includegraphics[width=\linewidth]{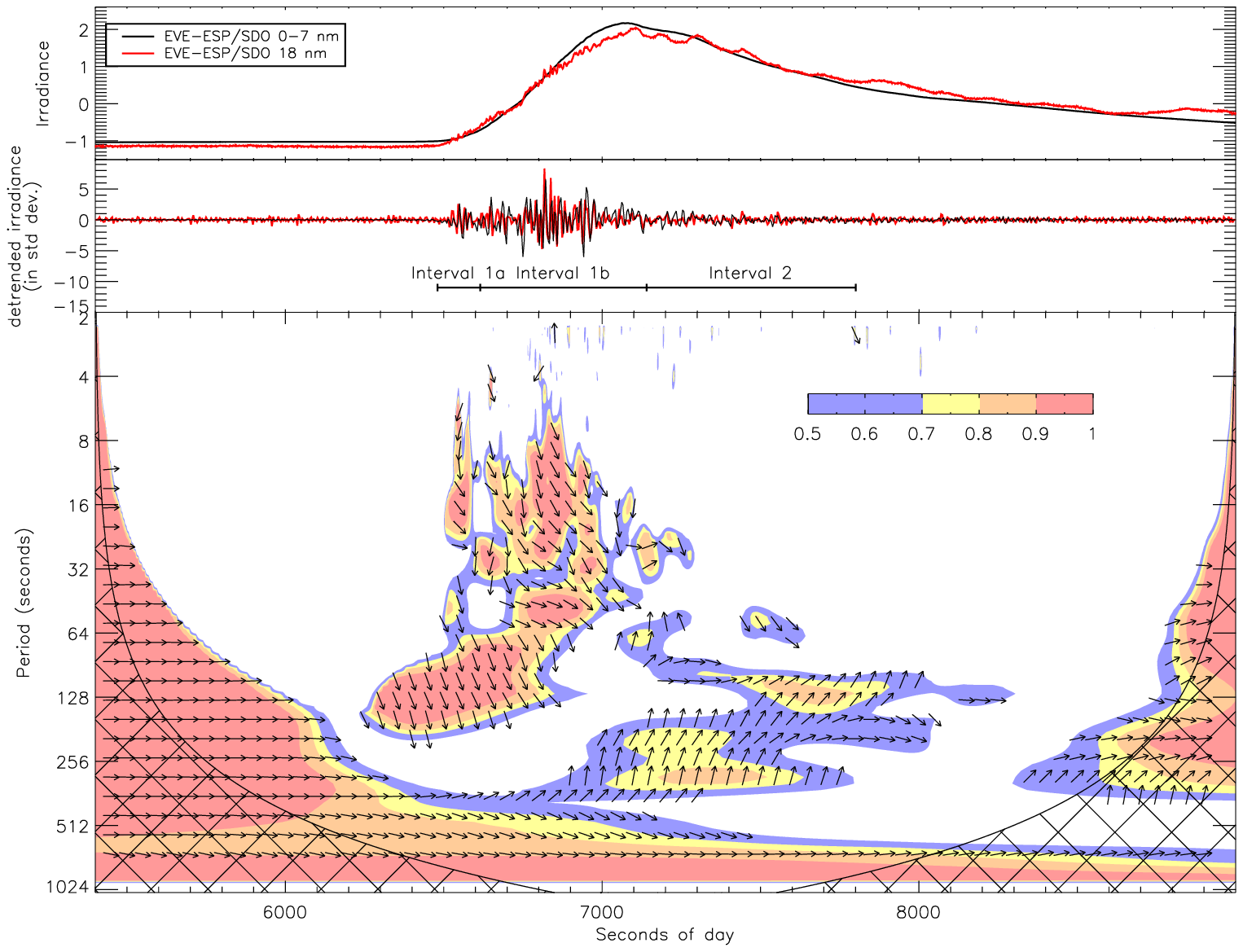}
	\end{center}
	\caption{Normalized light curves of the SXR and EUV 18~nm channels of SDO/EVE/ESP during the flare (top) and detrended signals (middle, for comparison only). The bottom panel shows wavelet coherence between both channels (without detrending). The overlaid arrows indicate the phase difference between the SXR and the EUV channels in the given time scale. They follow the trigonometric usage: $0$ (or ``in-phase'') and $\pi/2$ (lag) correspond to arrows pointing to the right and the top, respectively. 	}
	\label{fig cross coherence}
\end{figure*}

This method used above disregards the frequency composition of the signal and, moreover, only responds to short time-scale fluctations. Computing the wavelet coherence between the signals, using the method of \citet{Torrence99}\footnote{software available at \url{http://atoc.colorado.edu/research/wavelets/}} may return additional information and can be performed without removing the long-term trend, but this method also has limitations. In particular, no useful information can be recovered for signals that are very similar on long time-scales, such a flare pulse, because all periodicities are found to be coherent and therefore cannot be distinguished. Nonetheless, it did prove useful for some of our data.

Figure~\ref{fig cross coherence} shows the wavelet coherence between the ESP SXR and 18~nm channels. The signals have been first normalized by subtracting the average and then dividing by the standard deviation (top panel). The detrended signals are only shown in the middle panel for comparison with the previous method. The bottom panel shows the correlation coefficient as a function of time and period. Arrows are overlaid to show the phase difference between the channels. The cross-hatched area correspond to the so-called ``cone of influence'', where results are influenced by edge effects. 

Not surprisingly, we find enhanced wavelet coherence during intervals~1a and 1b for periods between 8 and 32~s.  This shows that those fluctuations are broad-band, like for most QPPs \citep{Nakariakov09}. We also find common oscillations during the impulsive phase in the 1--2~minutes range (lasting at least 2 periods). 
In both ranges of periods, the phase delay is close to $-\pi/2$; that is to say the EUV is leading the SXR channel by about a quarter of a period. 
During interval~2, we find common oscillations with periods of 3--5~minutes. This time, it is the EUV channel that lags the SXR channel up to  $\pi/2$. This range of periods is reminiscent of the \emph{p}-modes.\footnote{\emph{P}-modes have been analyzed with the ESP SXR channel by \citet{Didkovsky11} for ``quiet'' (nonflaring) time periods.}

We also find lagging and leading behaviors in wavelet coherence for other channels during intervals~1a and 1b (not shown). In particular, the RHESSI 3--6, 6--12, 12--25, and 25--50~keV channels show matching wavelet power in the 1--2~minute range with the ESP SXR channel, and all lead this channel by about $\pi/2$. As a cross-verification, the wavelet coherence analysis shows that they are nearly in phase with the ESP 18~nm channel in the 1--2~minute range. 

The Nobeyama 3, 9 and 17~GHz radio channels also show matching wave power with the ESP SXR and 18~nm channels in the 1--2~minute ranges; they are nearly in phase with ESP SXR, but lag ESP 18~nm by about $-\pi/2$. For the 8--32~s-period range, their phase is different during intervals~1a and 1b ($\approx -\pi/2$ and $\approx 0$, respectively), but consistent with the analysis of the time lags presented in Figure~\ref{fig time lag}.

\begin{figure*}
	\begin{center}
	\includegraphics[width=\linewidth]{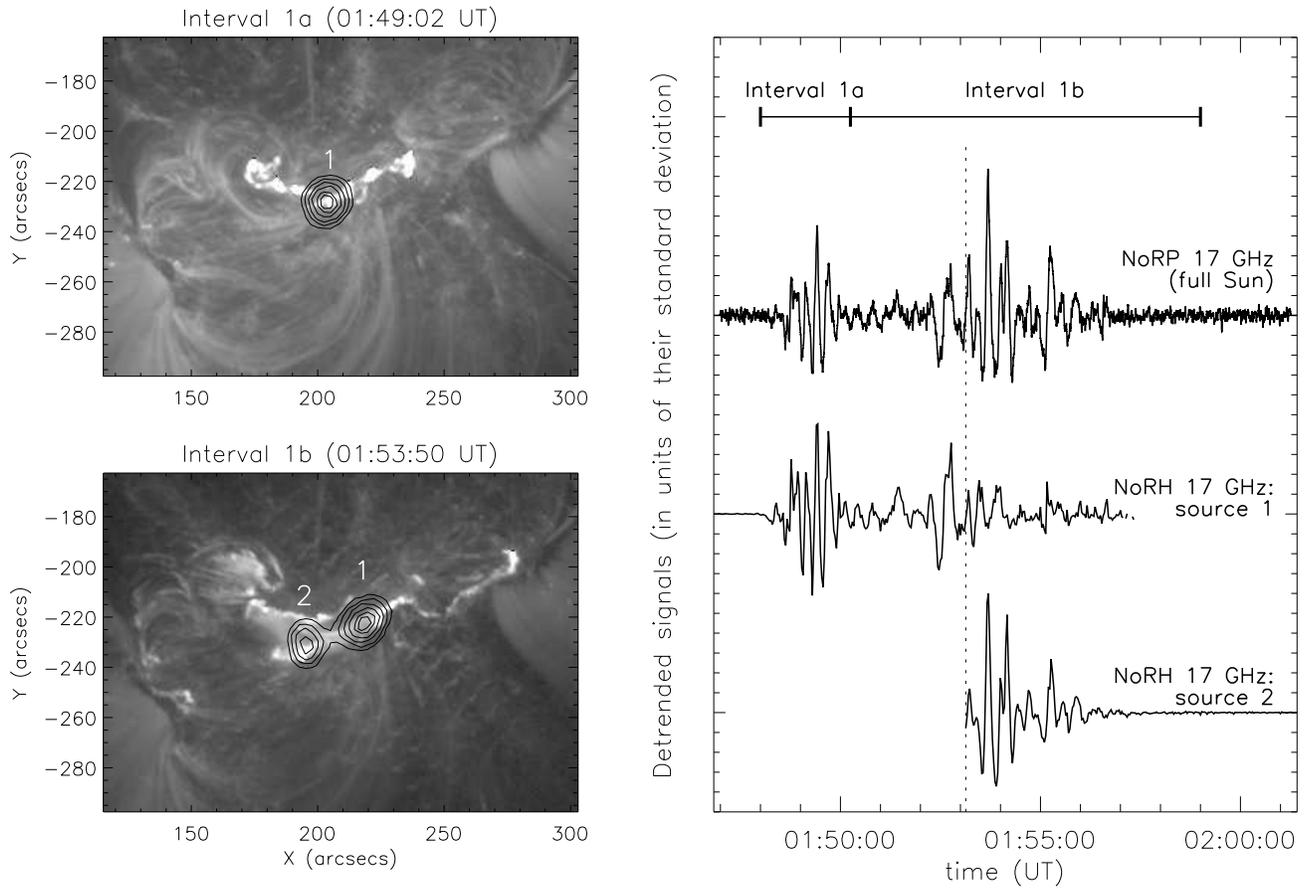}
	\end{center}
	\caption{AIA 171~\AA{} images overlaid with NoRH~17 GHz contours  at 95, 90, 80, 70, 60 and 50\% of maximum (left panels) and NoRP (full Sun) and NoRH detrended light curves for sources labeled ``1'' and ``2'' (right panel). 
}
	\label{fig pos sources}
\end{figure*}

To determine the spatial localization of the QPPs emission, one must use images of the solar atmosphere. 
Unfortunately, in most of the bandpasses (e.g. SDO/AIA, \emph{Hinode}/XRT, RHESSI) the cadence of the image acquisition is too low. The only exception is the NoRH data at 17~GHz, which have 2~s cadence, sufficient to detect the sources of QPP.
Figure~\ref{fig pos sources} shows that a radio source was present during interval~1a (upper left panel) and progressively moved to the West at 01:51. It progressively faded, but was still visible during interval~1b (lower left panel). Such motion is not unusual, e.g. \citet{Zimovets09}. 
A second source started to be resolved at 01:53:08; before, both sources overlapped within the instrumental resolution. From the comparison of the detrended light curves of both sources and that of the full Sun, we conclude that the fluctuations during both intervals~1a and 1b are then emitted by distinct structures. 

%
%
%
%
%
\begin{deluxetable*}{lrrccc}
\tablecolumns{6}
\tablewidth{0pc}
\tablecaption{Observations \label{tab observations}}
\tablehead{
\colhead{} & \colhead{Time Cadence}   & \colhead{Fluct.\tablenotemark{a}} & \multicolumn{3}{c}{Correlation coefficient\tablenotemark{b}} \\
\cline{4-6} \\
\colhead{Channel} & \colhead{(ms)}   & \colhead{(\%)} & \colhead{Ph. 1a} & \colhead{Ph. 1b} & \colhead{Ph. 2} \\
}
\startdata
PROBA2/LYRA Zr ($<2$~nm + 6--20~nm)  &    50 &   0.7 & 0.97 & 0.95 & 0.41 \\
PROBA2/LYRA Al ($<5$~nm + 17--80~nm) &    50 &   0.8 & 0.98 & 0.95 & 0.41 \\
NoRP 1~GHz                            &   100 &  20.8 & 0.002 & 0.07 & 0.13 \\
NoRP 2~GHz                            &   100 &  17.1 & 0.24 & 0.14 & 0.08 \\
NoRP 3~GHz                            &   100 &   7.0 & 0.45 & 0.32 & 0.13 \\
NoRP 9~GHz                            &   100 &   8.4 & 0.53 & 0.57 & 0.06 \\
NoRP 17~GHz                           &   100 &  14.8 & 0.63 & 0.55 & 0.04 \\
NoRP 35~GHz                           &   100 &   9.9 & 0.26 & 0.25 & 0.04 \\
SDO/EVE/ESP soft X-rays (0--7~nm)     &   250 &   0.4 & n/a  & n/a  & n/a \\
SDO/EVE/ESP 18~nm (17.2--20.6~nm)    &   250 &   5.7 & 0.75 & 0.74 & 0.12 \\
SDO/EVE/ESP 26~nm (23.1--27.6~nm)    &   250 &   5.5 & 0.58 & 0.67 & 0.08 \\
SDO/EVE/ESP 30~nm (28.0--31.6~nm)    &   250 &   6.6 & 0.74 & 0.74 & 0.14 \\
GOES ``short'' (0.05--0.4~nm)        &  2000 &   0.5 & 0.83 & 0.77 & 0.46 \\
GOES ``long'' (0.1--0.8~nm)          &  2000 &   0.3 & 0.95 & 0.90 & 0.55 \\
RHESSI 3--6~keV                      &  4000 &   7.0 & 0.29 & 0.23 & 0.11 \\
RHESSI 6--12~keV                     &  4000 &  12.8 & 0.22 & 0.14 & 0.10 \\
RHESSI 12--25~keV                    &  4000 &  11.0 & 0.44 & 0.23 & 0.15 \\
RHESSI 25--50~keV                    &  4000 &  15.6 & 0.57 & 0.55 & 0.11 \\
RHESSI 50--100~keV                   &  4000 &   7.2 & 0.51 & 0.52 & 0.07 \\
RHESSI 100--300~keV                  &  4000 &  20.5 & 0.34 & 0.21 & 0.11 \\
\enddata
\tablenotetext{a}{Maximum amplitude of fluctuations on time-scales shorter than 20~s (Figure~\ref{fig light curves and detrended}), relative to the increase of signal during the flare. }
\tablenotetext{b}{Correlation with SDO/EVE/ESP SXR channel; we report the value corresponding to the best time lags shown in Figure~\ref{fig time lag}. }
\end{deluxetable*}
%
\section{Discussion and Conclusions}
%
We analyzed fluctuations on time scales $\approx$~8--32~s during the X2.2 flare of 15~February~2011. Their fine structure was similar in EUV, radio, SXR and HXR channels during the impulsive phase, but with time delays up to $\approx 9$~s. During the declining phase, the presence of fluctuations is less certain; they are only visible in the SXR channels, but still with time delays between channels. 

There have been many studies of time delays in fluctuations during flares; for example between radio wavebands and HXR \citep[e.g.][]{Kaufmann83, Cornell84} or between HXR channels of different energy \citep{Aschwanden97}.  To our knowledge, this is the first time that EUV fluctuations leading SXR fluctuations have been observed on these time scales. We note, however, that \citet{Emslie78} measured lags $\lesssim 5$~s in EUV radiation relative to HXR for pulses on similar time scales as those we analyze. 

Time shifts between clocks are of concern, but do not affect our general conclusions because there are significant time delays between channels of the same instrument, like the EUV and SXR channels of EVE/ESP. 

A key question is the emission mechanism at work in the different wavebands. There is no doubt that HXR are emitted through bremsstrahlung (thermal or most probably non-thermal). But in the absence of observations with both high spectral and time resolution ($\approx 1$~s), 
we have no evidence of whether the SXR or EUV fluctuations occur in  emission lines (through collisional excitation of the ions, mainly by thermal electrons) or in the continuum (implying bremsstrahlung). 

Three physical mechanisms can produce time delays in the range of HXR energies \citep{Aschwanden04, Holman11}: time-of-flight (TOF) dispersion of free-streaming electrons, magnetic trapping with the collisional precipitation of electrons, and cooling of the thermal plasma \citep[Neupert effect,][]{Neupert68}. 

Electrons of higher energy have smaller TOF from loop top to the chromosphere, where they typically produce hard X-ray photons. Therefore, the TOF effect introduces 10--100~ms delays in the 25--50~keV light curves as compared to 50--100~keV \citep[][]{Aschwanden95}. 
As a consequence, TOF delays cannot explain our observations: first, we observe larger delays than expected between the HXR channels ($\approx 1$~s); second, the 50--100~keV channel should lead the 25--50~keV channel during interval~1b. 

Due to magnetic mirroring between the loop footpoints, electrons of higher energy precipitate later because they are less likely to escape the magnetic trap. 
\citet{Aschwanden97} finds typical delays of 1--10~s between HXR pulses at 200~keV and 50~keV and model them with trapping effects. The $\approx 1$~s delay between the two HXR channels during interval~1b is compatible with their results. The mismatch in the order of the channels in interval~1a could be explained by the effect of uncertainties, especially as this interval is shorter and contains fewer peaks for cross-correlation. 

Radio signals at 17 and 9~GHz are delayed with respect to the 25--50~keV signal by about 2~s during intervals~1a and 1b. This is expected for trapped highly relativistic electrons (100--1000~keV) that produce gyro-synchrotron emission in those wavebands \citep{Dulk85}. 
The variation in delay from interval~1a to 1b shown by HXR and radio channels coincides with  different spatial locations of the radio sources (Figure \ref{fig pos sources}). It is likely that the physical conditions (loop length, magnetic field) changed so that trapping times differ during those time intervals. As explained above, results in interval~1a must be interpreted with caution, though. 

Cooling effects introduce delays in the peak times of channels more dominated by thermal effects (e.g. SXR) with respect to channels that present a more impulsive behavior \citep[microwaves, HXR; e.g.][]{Aschwanden07, Jain11}. Cooling effects are compatible with short-period fluctuations in SXR channels lagging those in the impulsive HXR channels, as we observe. 
However, the same cooling effects are responsible for the flare peak to occur first in the GOES 0.5--4~\AA{} then in the GOES 1--8~\AA{} and in ESP 0--7~nm channels (see Figure~\ref{fig light curves and detrended}). Paradoxically, we observe exactly the reverse order in the short time scales fluctuations (Figure~\ref{fig time lag}). 

\citet{Dennis85} reports on nearly simultaneous ($<1$~s) pulses in UV and HXR, as a result of chromospheric heating by precipitating electrons. We suppose that the EUV fluctuations during intervals~1a and 1b also originate from chromospheric heating, which explains the observed simultaneity with HXR channels ($\lesssim 1$~s during interval~1b). 
 
Can the observed fluctuations be modulated by MHD waves? 
\citet{Vandoorsselaere11} used PROBA2/LYRA to observe oscillations with two distinct periods during a flare: both $\approx 8.5$~s and 
$\approx 75$~s period oscillations appear in the Zr, Al and Lyman~$\alpha$ channels. They interpreted the shorter periods in terms of the standing fast sausage mode and the longer periods in terms of the standing slow sausage mode. It is tempting to interpret our observations similarly as standing fast and slow sausage modes for the shorter (8--32~s) and longer (1--2~minute) periods respectively. 
However, apart from the similarity with the mode periods, our observations cannot confirm or invalidate the modulation of the electron beam by MHD waves. Besides, MHD waves alone can hardly explain the observed delays between the signals in HXR, EUV and SXR channels.

In summary, we suggest that the fluctuations observed during the impulsive phase (intervals~1a and 1b) were produced by a beam of precipitating electrons, possibly modulated by MHD waves. When electrons reached the denser layers, they excited EUV as well as HXR emission in lower energy channels (12--25 and 20--50~keV). 
Higher energy electrons were delayed by trapping effects: the higher the energy, the longer the delay to escape the trap. Therefore, the fluctuations appeared a few seconds later in 50--100~keV HXR and then in 9 and 17~GHz radio channels. 
 
The case of the SXR channels is more puzzling.  
The delays in these channels compared to EUV and HXR channels could be due to the Neupert effect, although the relative delays between the SXR channels are in reverse order than what is expected in this case. 
This point certainly deserves more investigation and modeling. 

\acknowledgments
%
We thank V. Nakariakov for useful discussions, the referee for his help in clarifying the discussion and the teams of GOES, NoRH, NoRP, PROBA2/LYRA, RHESSI, SDO/EVE and SDO/AIA for their open data use policy. Wavelet software was provided by C.~Torrence and G.~Compo.  LYRA is a project of the Centre Spatial de Liege, the Physikalisch-Meteorologisches Observatorium Davos and the Royal Observatory of Belgium funded by the Belgian Federal Science Policy Office (BELSPO) and by the Swiss Bundesamt f\"ur Bildung und Wissenschaft.  

{\it Facilities:} \facility{GOES}, \facility{NoRH}, \facility{NoRP}, \facility{PROBA2 (LYRA)}, \facility{RHESSI}, \facility{SDO (EVE)},  \facility{SDO (AIA)}
%

\end{document}